\documentclass[twocolumn,amsmath,amssymb,prd]{revtex4}

\def\tr{\mathrm{tr}}

\newcommand{\Slash}[1]{\!\!\not\!{#1}}
\newcommand{\SLASH}[1]{\not\!\!{#1}}
\newcommand{\half}{{\textstyle{1\over 2}}}

\usepackage{graphicx}
\newcommand{\be}{\begin{equation}}
\newcommand{\ee}{\end{equation}}
\newcommand{\ben}{\begin{eqnarray}}
\newcommand{\een}{\end{eqnarray}}
\def\pls{\partial\!\!\!/}
\def\bs{b\!\!\!/}
\def\ps{p\!\!\!/}
\def\As{A\!\!\!/}

\def\g{\gamma}

\def\m{\mu}

\def\bb{\bibitem}
\def\half{{\textstyle{1\over 2}}}

\newcommand{\incps}[5]{\includegraphics[#2,#3][#4,#5]{#1}}

\newcommand{\remark}[1]{}
\newcommand{\fgl}[1]{\hspace{0.75cm}#1\hspace{-0.75cm}}

\begin{document}

\title{Remarks on Lorentz and CPT violation in field theory}

\author{T. Mariz, J.R. Nascimento
and E. Passos}
\affiliation{Departamento de F\'\i sica, Universidade Federal da Para\'\i ba, \\
Caixa Postal 5008, 58051-970 Jo\~ao Pessoa, Para\'\i ba, Brazil}

\email{tiago,jroberto,passos@fisica.ufpb.br}

\date{\today}

\begin{abstract}
In this brief review we explicitly calculate the radiative corrections to the Chern-Simons-like term in the cases of zero and finite temperature, and in the gravity theory. Our results are obtained under the general guidance of dimensional regularization.
\end{abstract}

\maketitle

\section{Introduction}

The possibility of breaking Lorentz and CPT symmetries has been considered in several different contexts \cite{cfj,ck,ck1,cg,cg1,k}. Most of them were dedicated to the extended quantum electrodynamics (QED) sector of the 
extended standard model (see e.g. Ref. \cite{ck,ck1}). The initial motivation for consideration of 
Lorentz violation came from string theory \cite{va,kost}. Its basic idea is that interactions in the underlying theory induce nonzero expectation values for one or more Lorentz tensors which can be regarded as background quantities in the vacuum throughout spacetime. 

Some time ago, in Ref. \cite{cfj} the authors formulated the first model in which the electrodynamics is modified by adding
the Chern-Simons-like term to Maxwell term and verified that the Lorentz and CPT symmetries are broken. This model predicts the rotation of the plane of polarization of light from distance galaxies,
an effect which was not observed yet \cite{cfj}. In recent papers \cite{JP, L},  the modification of general relativity obtained by adding the Chern-Simons-like gravitational term has been studied. 
The authors have observed that in such modified theory the Schwarzschild metric turns out to be a
solution, gravitational waves possess two polarizations which travel with 
the velocity of light and polarized waves are suppressed or enhanced. 

In four dimensions Colladay and Kosteleck\'y posed the question whether such a 
term is induced when Lorentz and CPT violating term $\bar\psi\bs\g_5\psi$ is 
added to conventional Lagrangian of the QED \cite{ck1}. 
In this case the fermions can be integrated out,
and the radiative correction up to the first loop may lead to
\be
S_{CS} = \frac{1}{2}\int d^4x\epsilon^{\mu\nu\lambda\rho}k_{\mu}F_{\nu\lambda}A_{\rho}\label{ACS}
\ee
where $k_{\mu} = Cb_{\mu}$. The issue has been carefully investigated in several different contexts and regularization schemes, by many authors \cite{ck1,2,3,4,5,6,7,7a,8,9,10,11,12,13,14}, leading to results where $C$ 
vanishes \cite{ck1,cg1,7a,8,14} and results where $C$ does not \cite{2,3,4,5,6,7,9,10,11,12,13} but different each other.
This dependence upon the regularization scheme \cite{2,3} corresponds to an ``ambiguity'', i.e. finite but undetermined values, which has been well discussed in the literature \cite{ja99,mpv}.

In this brief review we will summarize the induction  of the Chern-Simons-like term by radiative corrections in the contexts of the zero and finite temperature, and in the gravity theory.

\section{Radiatively induced Chern-Simons-like term}

In order to study this issue, one may consider the action of the Lorentz and CPT violating fermion sector
\begin{equation}
S_f = \int d^4x \bar\psi(i\pls - m - \gamma_5 \bs -\As)\psi,
\end{equation}
where $b_{\mu}$ is a constant four vector which selects a fixed direction in spacetime. 
To account for the fermionic
integration we write
\be
e^{i S_{eff}[b,A]}=\int D\bar\psi D\psi\, e^{iS_f}
\ee
where the effective action is given by
\be\label{eff}
S_{eff}[b,A]=-i\,{\rm Tr}\,\ln(i\pls- m - \gamma_5 \bs -\As)\psi.
\ee
Note that the Eq. above can be written as $S_{eff}[b,A] = S_{eff}^{(0)}[b] + S_{eff}^{(1)}[b,A]$, where $S_{eff}^{(0)}[b]=-i\,{\rm Tr}\ln(\ps-m-\gamma_5\bs)$ and
\be\label{ea}
S_{eff}^{(1)}[b,A]=i\,{\rm Tr} \sum_{n=1}^{\infty}\frac1n
\Biggl[\frac1{\ps-m-\gamma_5\bs}{\As}\Biggr]^n.
\ee
Since the term $S_{eff}^{(0)}[b]$ is independent of the gauge field and cannot induce Chern-Simons-term, we shall focus only on the second term $S_{eff}^{(1)}[b,A]$ looking for the first order derivative terms which are linear in $\,\Slash{b}$ and quadratic in $\SLASH{A}$. This is the perturbative route, in which we do not include the contribution involving the vector $b_{\mu}$ into the Dirac propagator (see Ref. \cite{2} for details). 

Using the derivative expansion \cite{de1,de5}
\be\label{iden}
\frac1{{/\!\!\!p}-i{/\!\!\!\partial}-m}=\frac1{{/\!\!\!p}-m}+
\frac1{{/\!\!\!p}-m}\;i{/\!\!\!\partial}\;\frac1{{/\!\!\!p}-m}+\cdots,
\ee
we can rewrite Eq.(\ref{ea}) in the form
\be\label{ea1}
S^{(1b)}_{eff}[b,A]=\frac{i}{2}\int d^4x\;\left(\Pi_1^{\mu\nu}+
\Pi_2^{\mu\nu}+\Pi_3^{\mu\nu}\right)\,A_{\mu}A_{\nu}
\ee
where
\be\label{pi1}
\Pi_{1}^{\mu\nu}={\rm tr}\int\frac{d^4p}{(2\pi)^4}\;S(p){/\!\!\!b}\gamma_5S(p)
\gamma^{\mu}S(p)i{/\!\!\!\partial}S(p)\gamma^{\nu},
\ee
\be\label{pi21}
\Pi_2^{\mu\nu}={\rm tr}\int\frac{d^4p}{(2\pi)^4}\;S(p)
\gamma^{\mu}S(p){/\!\!\!b}\gamma_5S(p)i{/\!\!\!\partial}S(p)\gamma^{\nu},
\ee
and
\be\label{pi22}
\Pi_3^{\mu\nu}={\rm tr}\int\frac{d^4p}{(2\pi)^4}\;S(p)\gamma^{\mu}
S(p)i{/\!\!\!\partial}S(p){/\!\!\!b}\gamma_5S(p)\gamma^{\nu}.
\ee
with $S(p)=(\,\Slash{p}-m)^{-1}$. 

We evaluate the integrals under the general guidance of dimensional
regularization \cite{dr1,dr2,dr3}. Thus, we change dimensions from
$4$ to $2w$, and we change $d^4p/(2\pi)^4$ to
$(\mu^2)^{2-w}[d^{2w}p/(2\pi)^{2w}]$, where $\mu$ is an arbitrary
parameter that identifies the mass scale. We use two distinct routes
to do the calculations involving the Dirac matrices. In the first route
we use the cyclic property of the trace, to move $\gamma_5$ to the very
end of every expression involving the trace of Dirac matrices and $\gamma^\alpha\gamma_\alpha=2w$, so that we get
\begin{eqnarray}\label{ifinal}
S^{(1b)}_{eff}[b,A]&=&\frac32\,i\,\Pi(w)\,{\rm tr}(\gamma^{\mu}
\gamma^{\nu}\gamma^{\lambda}\gamma^{\rho}\gamma_5) \nonumber\\
 && \times \,b_{\mu}\int d^4x\,
\partial_{\nu}A_{\lambda}A_{\rho}.
\end{eqnarray}
Here the factor $3$ accounts for identical contributions
that comes from $\Pi_1^{\mu\nu}$, $\Pi_2^{\mu\nu}$ and
$\Pi_3^{\mu\nu}$, and $\Pi(w)$ is given by
\begin{eqnarray}\label{piw}
\Pi(w)&=&-\frac{2w-1}{96\pi^2}+\frac{w+1}{96\pi^2}\left(\frac{4\pi\mu^2}{m^2}
\right)^{2-w} \nonumber\\
&& \times\,\Gamma(2-w)(2-w).
\end{eqnarray}

In the above calculations we have set
$\Pi^{\mu\nu}_{i}=\Pi^{\mu\nu}_{i,{\rm div}}+\Pi^{\mu\nu}_{i,{\rm fin}}$
to split the $\Pi^{\mu\nu}_i$ contribution into two parts, one divergent
and the other finite. The contribution $\Pi_{i,{\rm div}}^{\mu\nu}$
is divergent in the limit $w\to2$, and it contribute with the term
proportional to $\Gamma(2-w)$. However, the factor involving the Dirac
matrices contributes with the term $(2-w)$, in a way such that the full
contribution is finite in the limit $w\to2$. Furthermore, this finite term
exactly compensates the finite contribution that appears from
$\Pi_{i,{\rm fin}}^{\mu\nu}$ in the limit $w\to2$. In the limit
$w\to2$ we can use 
${\rm tr}(\gamma^{\mu}\gamma^{\nu}\gamma^{\lambda}\gamma^{\rho}\gamma_5)
=4i\varepsilon^{\mu\nu\lambda\rho}$, but $\Pi(w\to2)\to0$ and this
leaves no room for Lorentz and CPT violation. The perfect balance between
the two contributions that we have just found has been identified before
in Ref.~{\cite{cgs}} as being peculiar to dimensional regularization. 

We stress that if one uses the relation $\{\gamma^\mu,\gamma_5\}=0$ to move
$\gamma_5$ to the end of every expression involving the trace of Dirac
matrices, the perfect balance between the two contributions is broken,
giving rise to a non zero value for the constant $C$. In the same way, if one uses the cyclic property of the trace and  $\gamma^\alpha\gamma_\alpha=4$ \cite{4}, we have
\be 
\label{k} k_{\mu} =
\frac{3}{16\pi^2}b_{\mu}. 
\ee 

\vspace*{1mm}

\noindent This is the unambiguous Chern-Simons coefficient \cite{2} obtained when we use the nonperturbative route, in which we include the contribution involving $\,\Slash{b}$ into the Dirac propagator. 

We make this point stronger by considering another route to implement the
calculation involving properties of the Dirac matrices when the spacetime
has dimension $2w$. We follow \cite{dr3,bm}, and now the Dirac matrices
contracted with $\Slash{b}$ and $\SLASH{A}$ are physical
matrices; they are written in the form ${\bar\gamma}^{\mu}$, etc. The other Dirac matrices are changed according to the rule
$\gamma^{\alpha}\to{\bar\gamma}^{\alpha}+{\hat\gamma}^{\alpha}$, where
$\{{\bar\gamma}^{\alpha},{\bar\gamma}^{\beta}\}=2{\bar g}^{\alpha\beta}$,
$\{{\hat\gamma}^{\alpha},{\hat\gamma}^{\beta}\}=2{\hat g}^{\alpha\beta}$,
and $\{{\bar\gamma}^{\alpha},{\hat\gamma}^{\beta}\}=0$, and also
${\bar\gamma}^{\alpha}{\bar\gamma}_{\alpha}=4$,
${\bar\gamma}^{\alpha}{\hat\gamma}_{\alpha}=0$ and
${\hat\gamma}^{\alpha}{\hat\gamma}_{\alpha}=2(w-2)$. In this case we can use either the cyclic property of the trace, or the
relations $\{\gamma_5,{\bar\gamma}^{\mu}\}=[\gamma_5,{\hat\gamma}^{\mu}]=0$
with ${\rm tr}({\bar\gamma}^{\mu}{\bar\gamma}^{\nu}
{\bar\gamma}^{\lambda}{\bar\gamma}^{\rho}\gamma_5)=4i
\varepsilon^{\mu\nu\lambda\rho}$ and ${\rm tr}({\gamma}^{\mu}{\gamma}^{\nu}
{\gamma}^{\lambda}{\hat\gamma}^{\rho}\gamma_5)=0$, that we arrive at the same result where the Eq. (\ref{piw}) does vanish.

Therefore, we can emphasize that concerning dimensional regularization scheme a variety of results can be obtained when a variety of prescriptions is made. 

\section{Radiatively induced Chern-Simons-like term at finite temperature}

In this section we will analyze the behavior of the parameter C when we take
temperature into account. By comparing with results in the literature \cite{nrs,cgs,ezr}, we also will find that at finite temperature, the Chern-Simons-like term remains undetermined. It is more convenient rewrite the Eq. (\ref{eff}) as $S_{eff}[b,A] = S_{eff}^{(0)}[b] + S_{eff}^{(1)}[b,A]$, where now
\begin{widetext}
\begin{eqnarray}\label{snew}
S_{eff}^{(1)}[b,A] = i\int_0^1\,dz\,{\rm Tr} \left[\frac{1}{i\pls - m - \gamma_5\bs -
z\As(x)}\As(x)\right].
\end{eqnarray}
To perform the momentum space integration in Eq. (\ref{snew}), we consider the prescription \cite{lhm}: \be
i\,\pls\to\ps,\qquad\qquad
\As(x)\to\As\left(x-i\frac{\partial}{\partial p}\right).\ee 
Then, the Eq. (\ref{snew}) now reads
\ben \label{seff1} S_{eff}^{(1)}[b,A]
&=& i \int_0^1 dz \int{d^4x}\int\frac{d^4p}{(2\pi)^4}\,{\rm
tr}\left[\frac{1}{\ps - m - \gamma_5\bs -
ze\As(x-i\frac{\partial}{\partial p})}\As(x)\right]. 
\een 
We can manipulate the Eq.(\ref{seff1}) to keep only first order derivative terms which are
linear in $\bs$ and quadratic in $\As$. Carrying out the integral in
$z$ gives 
\ben \label{seff2} S_{eff}^{(1b)}[b,A] &=&
-\frac{i}{2}\int d^4 x \int \frac{d^4 p}{(2 \pi)^4} {\rm tr} \left[\frac{1}{\ps - m}i\partial_{\m}\As
 \frac{\partial}{\partial p_{\m}}\frac{1}{\ps - m}\gamma_5
 \bs\frac{1}{\ps - m}\As + \frac{1}{\ps - m}\gamma_5\bs \frac{1}{\ps - m}i
\partial_{\m}\As\frac{\partial}{\partial p_{\m}}\frac{1}{\ps
- m}\As \right].
\een 
\end{widetext}
Now, using the relation
$$
\frac{\partial}{\partial p_\mu}\frac{1}{\ps -m} = -\frac{1}{\ps -m}
\gamma^{\mu}\frac{1}{\ps-m}.
$$
and taking the traces of the products of $\gamma$ matrices on
relevant terms, i.e., the terms that contain
$\tr(\gamma^{\mu}\gamma^{\nu}\gamma^{\alpha}\gamma^{\beta}\gamma_5)$,
the  Eq.(\ref{seff2}) takes the form \be \label{seff3}
S_{eff}^{(1b)}[b,A] = -\frac{1}{2}\int d^4\, x\int \frac{d^4\,
p}{(2 \pi)^4} \frac{N}{(p^2 - m^2)^4}, \ee 
where $N$ is given by
\ben \label{numer} N &=& -4i(p^2 -
m^2)\left[\epsilon^{\alpha\beta\m\sigma}\left(3m^2+p^2\right)-4
\epsilon^{\alpha\beta\m\nu} p_{\nu} p^{\sigma}\right] \nonumber\\
&\times&b_\sigma
\partial_{\mu}A_\alpha A_\beta.
\een Note that by power counting the momentum integral in
Eq.~(\ref{seff3}) contains terms with logarithmic divergence. Let us use
the relation
\be\label{re}
\int\frac{d^Dq}{(2\pi)^D}q_{\mu}q_{\nu}f(q^2)=\frac{g_{\mu\nu}}{D}
\int\frac{d^Dq}{(2\pi)^D}q^2f(q^2), 
\ee 
that naturally removes the logarithmic divergence. Now, considering $D=4$, the terms containing
$p^2$ and $p_{\nu}p^{\sigma}$ in (\ref{numer}) cancel out and we
find \be N = -12m^2i(p^2 - m^2)\epsilon^{\alpha\beta\m\sigma}
b_\sigma
\partial_{\mu}A_\alpha A_\beta.
\ee  In this way, the logarithmic divergence in (\ref{seff3})
disappears, so that the effective action now reads \ben
\label{seff33} S_{eff}^{(1b)}[b,A]& =&\left[ {6im^2}\int
\frac{d^4\, p}{(2 \pi)^4} \frac{1}{(p^2 - m^2)^3}\,\right] \\
\nonumber & \times & \epsilon^{\alpha\beta\m\sigma}{b_{\sigma}}\int
d^4\,x
\partial_{\mu}A_\alpha A_\beta,
\een which is finite by power counting. Evaluating the momentum
integral in the (\ref{seff33}) we also obtain unambiguously the
Chern-Simons coefficient \cite{2} 
\be \label{k} k_{\mu} =
\frac{3}{16\pi^2}b_{\mu}. \ee 
However, if we use another regularization scheme $k_{\mu}$ may vanish, for instance, in
Pauli-Villars regularization scheme \cite{ck1}. Next we study such undetermined coefficient
when we take  into account the temperature.

\begin{figure}[h]
\centerline{\includegraphics[{angle=90,height=7.0cm,angle=180,width=8.0cm}]
{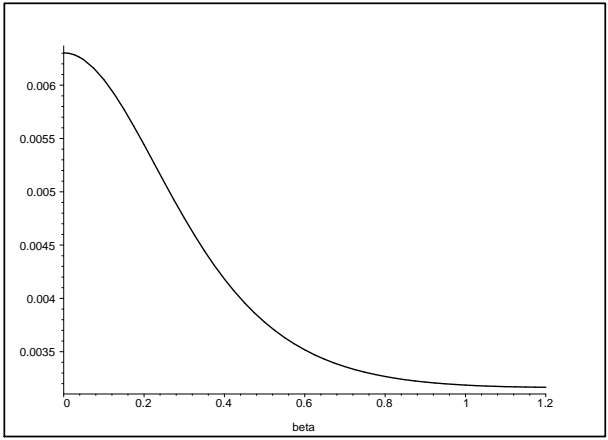}} \caption{The function $f(m^2, \beta)$ is different from
zero everywhere. At zero temperature ($\beta\to\infty$), the
function tends to a nonzero value ${1}/{32\pi^2}$.}\label{fig2}
\end{figure}

Let us now assume that the system is at thermal equilibrium with a
temperature $T=1/{\beta}$. In this case we can use Matsubara
formalism for fermions, which consists in taking $p_0 = (n +
1/2)2\pi/{\beta}$ and changing $(1/2\pi)\int dp_0=1/\beta\sum_n$
\cite{DJ}. We also change the Minkowski space to Euclidean space, by
making $x_0 = -ix_4$, $p_0=ip_4$ and $b_0=ib_4$, such that
$p^2=-p_E^2$, $p_E^2 = {\bf p}^2 + p_4^2$, $d^4p=id^4p_E$  and
$d^4x=-id^4x_E$. Now the Eq. (\ref{seff33}) can be written as 
\ben
\label{seff5} S_{eff}^{(1b)}[b,A] = 6 f(m^2, \beta)
\epsilon^{\alpha\beta\m\sigma}{b_{\sigma}} \int (-i)d^4\,x_E
\partial_{\mu}A_\alpha A_\beta,\;\;
\een 
where $f(m^2, \beta)$, is the Chern-Simons coefficient
dependent on the temperature which is given by \ben \label{f} f(m^2,
\beta)&=&\frac{m^2}{\beta}\int\;\frac{d^3{\bf p}
}{(2\pi)^3}\sum_{n=-\infty}^{\infty}
\frac{1}{({\bf p}^2+p_4^2 + m^2)^3}\\
&=&\nonumber\frac{m^2}{2\beta}\frac{d^2}{d(m^2)^2}\int\;\frac{d^3{\bf p}
}{(2\pi)^3}\sum_{n=-\infty}^{\infty}\frac{1}{({\bf p}^2+p_4^2 +
m^2)^2}.
\een 
We calculate the momentum integral by adopting
dimensional regularization scheme to obtain \be \label{t} f(m^2,
\beta) = \frac{m^2}{2\beta}\frac{\Gamma(3-D/2)}{(4\pi)^{D/2}}
\sum_{n=-\infty}^{\infty} \frac{1}{(p_4^2 + m^2)^{3-D/2}}. \ee 
To perform summation we shall use an explicit representation for
the sum over the Matsubara frequencies \cite{FO}: 
\ben\label{fo}
&& \sum_n [(n+b)^2 + a^2]^{-\lambda}= \frac{\sqrt{\pi}\Gamma(\lambda
- 1/2)}{\Gamma(\lambda)(a^2)^{\lambda - 1/2}} \\
&+& \nonumber 4\sin(\pi\lambda)\int_{|a|}^\infty \frac{dz}{(z^2 - a^2)^{\lambda}}
Re\left(\frac{1}{\exp 2\pi(z + ib) -1}\right), 
\een 
which is valid for $1/2<\lambda<1$. This implies that for $\lambda=3-D/2$ as given
in Eq.(\ref{t}) we cannot apply this relation for $D=3$, because the
integral in (\ref{fo}) does not converge. Thus, let us perform the
analytic continuation of this relation, so we obtain 
\begin{widetext}
\ben
\int_{|a|}^\infty \frac{dz}{(z^2 - a^2)^{\lambda}}
Re\left(\frac{1}{\exp 2\pi(z + ib) -1}\right) &=& \frac{1}{2a^2}\frac{3-2\lambda}{1-\lambda} \int_{|a|}^\infty \frac{dz}{(z^2 - a^2)^{\lambda-1}} Re\left(\frac{1}{\exp 2\pi(z + ib) -1}\right) \\
&& - \nonumber \frac{1}{4a^2}\frac{1}{(2-\lambda)(1-\lambda)} \int_{|a|}^\infty \frac{dz}{(z^2 - a^2)^{\lambda-2}} \frac{d^2}{dz^2}Re\left(\frac{1}{\exp 2\pi(z + ib) -1}\right). 
\een
\end{widetext}
Now for $D=3$ the Eq.(\ref{t}) takes the form \cite{mnprb} 
\be f(m^2, \beta) =
\frac{1}{32\pi^2}+\frac{1}{16}F(\xi), \ee where $\xi=\frac{\beta
m}{2\pi}$ and the function \be F(\xi)=
\int_{|\xi|}^{\infty}dz(z^2-\xi^2)^{1/2} \frac{\tanh(\pi
z)}{\cosh^2(\pi z)}, \ee approaches the limits:
$F(\xi\to\infty)\to0$ ($T\to0$) and $F(\xi\to0)\to{1}/{2\pi^2}$
($T\to\infty$) --- see Fig.\ref{fig2}. Thus, we see that at high
temperature the Chern-Simons coefficient is twice its value at zero
temperature, i.e., $f(m^2,\beta\to0)={1}/{16\pi^2}$. On the other
hand, at zero temperature, one recovers the result (\ref{k}).

\section{Radiatively induced Chern-Simons-like term in general relativity}

The action that we are interested is given by
\begin{equation}\label{S1}
S = \int\mathrm{d}^4x (\half i e e^\mu_{\,\,\,\,a} \bar\psi\gamma^a\stackrel{\leftrightarrow}{D}_\mu\psi-e e^\mu_{\,\,\,\,a}\bar\psi b_\mu \gamma^a\gamma_5\psi),
\end{equation}
where we have included the parity-violating term. Here, $e^\mu_{\,\,\,\,a}$ is the tetrad 
(vierbein), $e\equiv\det e^\mu_{\,\,\,\,a}$ and $b_\mu$ is a constant 4-vector. The covariant derivative is given by
\be
D_\mu\psi = \partial_\mu\psi + \half w_{\mu cd}\sigma^{cd}\psi,
\ee
where $w_\mu^{\,\,\,\,cd}$ is the spin connection and $\sigma^{cd} = \frac14[\gamma^c, \gamma^d]$, whereas the covariant derivative on a Dirac-conjugate field $\bar\psi$ is
\be
D_\mu\bar\psi = \partial_\mu\bar\psi - \half w_{\mu cd}\bar\psi\sigma^{cd}.
\ee
Using the expressions above we can rewrite the Eq. (\ref{S1}) as follow
\ben
S &=& \int\mathrm{d}^4x (\half i e e^\mu_{\,\,\,\,a} \bar\psi\gamma^a\stackrel{\leftrightarrow}{\partial}_\mu\psi + {\textstyle{1\over 4}} i e e^\mu_{\,\,\,\,a}  \bar\psi w_{\mu cd}\Gamma^{acd}\psi \nonumber\\
&& - e e^\mu_{\,\,\,\,a}\bar\psi b_\mu \gamma^a\gamma_5\psi),
\een
where $\Gamma^{acd}=\frac16(\gamma^a\gamma^c\gamma^d \pm permutations)$, i.e. the antisymmetrized product of three $\gamma$-matrices.

In the weak field approximation we consider 
$g_{\mu\nu}= \eta_{\mu\nu} + h_{\mu\nu}$ ($g^{\mu\nu}= \eta^{\mu\nu} - h^{\mu\nu}$),
which induces an expansion for the vierbein 
$e_{\mu a}= \eta_{\mu a} + \frac{1}{2}h_{\mu a}$ ($e^\mu_{\,\,\,\,a}= 
\eta^\mu_{\,\,\,\,a} - \frac{1}{2}h^\mu_{\,\,\,\,a}$). Then, the linearized 
Chern-Simons-like action takes the form \cite{JP} 
\be \label{slinear}
S_{linear}= \frac14\int\:d^4x h^{\mu\nu}v^\lambda\epsilon_{\alpha\mu\lambda\rho}\partial^\rho(\partial_\gamma\partial^\gamma h_\nu^\alpha-\partial_\nu\partial_\gamma h^{\gamma\alpha}).
\ee
The main goal here is to induce this action by radiative correction
of fermionic matter field obtaining the relation between $v_{\lambda}$ and $b_{\mu}$ \cite{mnpr}.
In order to perform this calculation we consider the fermionic model represented by the action 
\be
e^{i\Gamma[h]} = \int \mathcal{D}\bar\psi \mathcal{D}\psi\, e^{iS[h, \bar\psi, \psi]},
\ee
where the linearized effective action is given by
\be\label{secferm}
S[h, \bar\psi, \psi]=\int\mathrm{d}^4x(\half i\bar\psi\Gamma^\mu\stackrel{\leftrightarrow}{\partial}_\mu\psi+\bar\psi h_{\mu\nu}\Gamma^{\mu\nu}\psi-\bar\psi b_\mu\gamma^\mu\gamma_5 \psi),
\ee
with $\Gamma^\mu=\gamma^\mu-\frac12 h^{\mu\nu}\gamma_\nu$ and 
$\Gamma^{\mu\nu}=\frac12 b^\mu\gamma^\nu\gamma_5-\frac{i}{16}(\partial_\rho h_{\alpha\beta})\eta^{\beta\nu}\Gamma^{\rho\mu\alpha}$. 
In this expression, we neglect the terms proportional to $h=\eta^{\mu\nu}h_{\mu\nu}$ 
because they do not contribute to generating of the Chern-Simons-like action.

The Feynman rules that we obtain from Eq.(\ref{secferm}) are:
\ben
\raisebox{-0.4cm}{\incps{ferm.eps}{-1.5cm}{-.5cm}{1.5cm}{.5cm}}
 &=& S(p)=\frac{i}{\ps-m} \nonumber\\
\raisebox{-0.4cm}{\incps{fermm.eps}{-1.5cm}{-.5cm}{1.5cm}{.5cm}}
&=&  - i \bs \gamma_5 \nonumber\\
\raisebox{-.5cm}{\incps{vert.eps}{-1.5cm}{0cm}{1.5cm}{1cm}} &=& -{\textstyle{i\over 4}}\gamma_\mu(2p+q)_\nu \nonumber\\
\raisebox{-.5cm}{\incps{vertgam.eps}{-1.5cm}{0cm}{1.5cm}{1cm}} 
&=& i\gamma_\mu b_\nu\gamma_5 \nonumber\\
\raisebox{-.5cm}{\incps{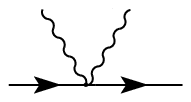}{-1.5cm}{0cm}{1.5cm}{1cm}} &=& -{\textstyle{i\over16}}\eta^{\beta\nu}\Gamma^{\mu\rho\alpha}(q_1-q_2)_\rho.
\een
\begin{figure}[h]
\centering
\begin{tabular}{ccccc}
\fgl{(a)}\includegraphics[width=0.15\textwidth]{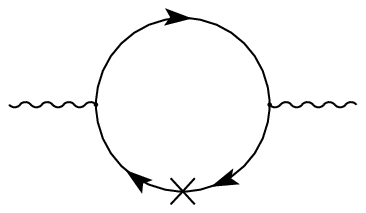} &
\fgl{(b)}\includegraphics[width=0.15\textwidth]{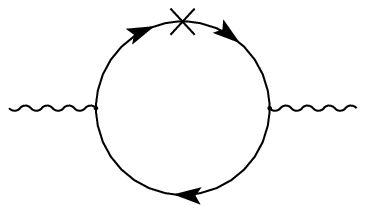} &
\end{tabular}
\caption{One-loop relevant contributions}
\label{oneloop}
\end{figure}
The relevant one-loop graphs to the Chern-Simons-like effective action are shown in the Fig. (\ref{oneloop}), whose Feynman integrals are given by
\begin{widetext}
\begin{equation}\label{pi1}
\Pi_{a}^{\mu\nu\alpha\beta}(q) = -\frac{i}{16}\,\mathrm{tr} \int\frac{\mathrm{d}^4p}{(2\pi)^4} \left[\gamma^\mu(2p+q)^\nu S(p)\gamma^\alpha(2p+q)^\beta S(p+q)\Slash{b}\gamma_5 S(p+q)\right]
\end{equation}
and
\begin{equation}\label{pi2}
\Pi_{b}^{\mu\nu\alpha\beta}(q) = -\frac{i}{16}\,\mathrm{tr} \int\frac{\mathrm{d}^4p}{(2\pi)^4} \left[\gamma^\mu(2p+q)^\nu S(p)\Slash{b}\gamma_5 S(p)\gamma^\alpha(2p+q)^\beta S(p+q)\right].
\end{equation}

It is straightforward to see that 
\be
\Pi^{\mu\nu\alpha\beta}(q)=\Pi_{a}^{\mu\nu\alpha\beta}(q)=\Pi_{b}^{\alpha\nu\mu\beta}(-q),
\ee
which appears of substituting the loop momenta, $p\rightarrow p-qx$, and we use the cyclic 
properties of the trace of a product of $\gamma$-matrices. 
So, from now on we work only with Eq. (\ref{pi1}) which takes the form
\begin{eqnarray}\label{pib1}
\Pi^{\mu\nu\alpha\beta}(q) &=& -\frac{1}{8}\int_0^1\mathrm{d}x\,x\, \int\frac{\mathrm{d}^4p}{(2\pi)^4} \frac{(2p+q(1-2x))^\nu (2p+q(1-2x))^\beta}{[p^2-m^2+x(1-x)q^2]^3} \nonumber\\
&&\times\mathrm{tr}\left[\gamma^\mu(\,\Slash{p}-\Slash{q}x+m)\gamma^\alpha(\,\Slash{p}+\Slash{q}(1-x)+m)\Slash{b}\gamma_5(\,\Slash{p}+\Slash{q}(1-x)+m)\right],
\end{eqnarray}
where we have used Feynman parameter to combine the denominator in  Eq. (\ref{pib1}). Taking into account the trace of Dirac matrices and dropping all odd terms in $p$, we get
\ben
\label{pi6}
\Pi^{\mu\nu\alpha\beta}(q)&=&-\frac{1}{8}\int_{0}^{1}dx\,x\int\frac{d^4p}{(2\pi)^{4}}
\frac{N^{\mu\nu\alpha\beta}(p^0, p^2, p^4)}{[p^{2}-m^{2}+x(1-x)q^{2}]^{3}},
\een
where the numerator $N^{\mu\nu\alpha\beta}(p^0, p^2, p^4)$ has the form,
\be
N^{\mu\nu\alpha\beta}(p^0, p^2, p^4)=4p^{\nu}p^{\beta}(T_{0}^{\alpha\mu}+T_{pp}^{\alpha\mu})+2(1-2x)(p^{\nu}q^{\beta}+p^{\beta}q^{\nu})(T_{p}^{\alpha\mu}+T_{ppp}^{\alpha\mu})+(1-2x)^{2}q^{\nu}q^{\beta}(T_{0}^{\alpha\mu}+T_{pp}^{\alpha\mu})
\ee
with
\ben
T_{0}^{\alpha\mu}&=&-4ib_{\lambda}\epsilon^{\alpha\mu\lambda\theta}q_{\theta}[x(1-x)^2q^2+(2-x)m^{2}],\\
T_{p}^{\alpha\mu}&=&-4ib_{\lambda}\epsilon^{\alpha\mu\lambda\rho}[m^{2}+(1-x^2)q^{2}]p_{\rho} - 8i(1-x)b_{\lambda}[\epsilon^{\mu\lambda\rho\theta}q^{\alpha}-
\epsilon^{\alpha\lambda\rho\theta}q^{\mu}-(1-x)\epsilon^{\alpha\mu\lambda\theta}q^{\rho}]q_{\theta}p_{\rho},\\
T_{pp}^{\alpha\mu}&=&-4ib_{\lambda}[2(\epsilon^{\mu\lambda\rho\theta}p_{\rho}p^{\alpha}-\epsilon^{\alpha\lambda\rho\theta}p_{\rho}p^{\mu}+x \epsilon^{\alpha\mu\lambda\rho}p_\rho p^\theta)-(2-x)\epsilon^{\alpha\mu\lambda\theta}p^{2}]q_{\theta},\\
T_{ppp}^{\alpha\mu}	&=&4ib_{\lambda}\epsilon^{\alpha\mu\lambda\rho}p^{2}p_{\rho}.
\een
\end{widetext}

The integral (\ref{pi6}) is badly divergent. Finally, using the dimensional regularization, the Eq. (\ref{pib1}) takes the form
\begin{equation}
\label{pib3}
\Pi^{\mu\nu\alpha\beta}(q) = b_\lambda\epsilon^{\alpha\mu\lambda\rho}q_\rho\left[A\,q^2 \eta^{\beta\nu} + 
B\,q^\beta q^\nu \right],
\end{equation}
where the expressions $A$ and $B$ are given by
\begin{widetext}
\begin{eqnarray}\label{pille}
A &=& \frac{1}{32\pi^2}\int_0^1\mathrm{d}x\,\left[3x^3(1-x)+(5x-3)x^2(1-x)\left(\frac{2}{\epsilon}+\ln\left(\frac{4\pi\mu^2}{-M^2}\right)-\gamma\right)\right. \nonumber\\ &&-\left.3x^2\left(\frac{2}{\epsilon}+\ln\left(\frac{4\pi\mu^2}{-M^2}\right)-\gamma+1\right)\frac{m^2}{q^2}\right]
\end{eqnarray}
and
\begin{eqnarray}
B &=& \frac{1}{64\pi^2}\int_0^1\mathrm{d}x\,\left\{(1-2x)^2(3-2x)\frac{x^2(1-x)q^2}{M^2}\right. \nonumber\\
&&+\left.\left[(2-3x)(1-2x)-4x(1-x)\right]x(1-2x)\left(\frac{2}{\epsilon}+\ln\left(\frac{4\pi\mu^2}{-M^2}\right)-\gamma\right)\right\},
\end{eqnarray}
\end{widetext}
after they have been expanded around $\epsilon \to 0$, with $\epsilon=4-D$ and $M^2=m^2-x(1-x)q^2$. 

As one can see $\int_0^1dx[(5x -3)x^2(1-x)](\frac{2}{\epsilon} - \gamma)=0$ in $A$ 
and $\int_0^1dx[(2-3x)(1-2x) - 4x(1-x)]x(1-2x)(\frac2{\epsilon}-\gamma)=0$ in 
$B$, then $A$ and $B$ take the form
\begin{eqnarray}
A &=& \frac{1}{32\pi^2}\int_0^1\mathrm
{d}x\,\left[3x^3(1-x)+\frac{(1-2x)x^3(1-x)^2q^2}{m^2-x(1-x)q^2}
\right. \nonumber \\ &-& \left.3x^2\left(\frac{2}{\epsilon}+\ln\left(\frac{4\pi\mu^2}{-M^2}\right)-\gamma+1\right)\frac{m^2}{q^2}\right]
\end{eqnarray}
and
\begin{eqnarray}
B &=& \frac{1}{32\pi^2}\int_0^1\mathrm{d}x\,\frac{(1-2x)^2x^2(1-x)q^2}{m^2-x(1-x)q^2}.
\end{eqnarray}
Here we have performed an integration by parts on $x$ for log term in $A$ and $B$.
Note that in $A$ the divergent part is present which will disappear when we consider
the limit $m^2\rightarrow 0$. Now performing the $x$-integration, we have
\begin{equation}\label{pi_general}
A|_{m^2\rightarrow 0}\, = - B|_{m^2\rightarrow 0}\, = \frac{1}{192\pi^2}.
\end{equation}
We substitute these results into Eq. (\ref{pib3}), to obtain the Chern-Simons-like term 
\begin{equation}
\Pi^{\mu\nu\alpha\beta}(q) = \frac{1}{192\pi^2}b_\lambda\epsilon^{\alpha\mu\lambda\rho}q_\rho\left[\,q^2 \eta^{\beta\nu} - 
\,q^\beta q^\nu \right].
\end{equation}
Finally, the Chern-Simons-like gravitational action induced by
radiatively corrections is given by
\be
\label{sf}
\Gamma_{\mathrm{cs}}[h] = \frac{1}{192\pi^2} \int\mathrm{d}^4x b^\lambda h^{\mu\nu}\epsilon_{\alpha\mu\lambda\rho}\partial^\rho\left[\partial_\gamma\partial^\gamma h^\alpha_\nu-\partial_\nu\partial_\gamma h^{\gamma\alpha}\right].
\ee
Comparing to Eq.(\ref{slinear}) we obtain the relation between the parameters 
$v_{\lambda}$ and $b_{\mu}$ which is written as 
\be
v_\lambda=\frac{1}{48\pi^2}b_\lambda.
\ee

\section{conclusion}

We have studied the induction of Chern-Simons-like term at zero and finite temperature, and in gravity theory. Here the dimensional regularization was applied to evaluate momentum integrals. We found that depending on the prescription, one can obtain either zero Chern-Simons coefficient or the unambiguous result \cite{2}, at zero temperature.

At finite temperature, our result is finite \cite{mnprb} but does not agree with other results presented in the literature \cite{nrs,cgs,ezr}. We argue that this is also caused by different regularization schemes. In the limit $T \rightarrow 0$ our result leads to a nonzero Chern-Simons-like term, a behavior also predicted in \cite{nrs}, obtained with the use of dimensional regularization, and the result in \cite{ezr}, obtained with the use of cutoff regularization scheme. However, it is in conflict with the result found in \cite{cgs} which
suggests the vanishing of the Chern-Simons-like term at zero temperature. On the other hand, at high temperature our result behaves as the result of \cite{cgs}. But now, however, it conflicts with the results in \cite{nrs,ezr} which predict that the Chern-Simons-like term vanishes at high temperature. These results are all finite, and they show that the Chern-Simons coefficient is indeed undetermined just as it happens at zero temperature \cite{ja99,mpv}. 

Finally, we recall that we also have calculated the radiative corrections induced by Dirac fermions coupled to a gravitational background field, including the nonstandard contribution that violates parity. In this calculation we have used the weak field approximation. As we are using the perturbative route, i.e., to lowest order in $b_\mu$, probably these radiative corrections are also undetermined.

\acknowledgments

This work was done in collaboration with D. Bazeia, R. F. Ribeiro and F. A. Brito. We would like to thank to A. Yu. Petrov 
for comments and discussions, and CAPES, CNPq, PADCT/CNPq, and PRONEX/CNPq/FAPESQ for financial support.

\end{document}